# Evaluating the Effectiveness of LLMs in Fixing Maintainability Issues in Real-World Projects

Henrique Nunes
*Federal University of Minas Gerais*
Belo Horizonte, Brazil
henrique.mg.bh@gmail.com

Eduardo Figueiredo
*Federal University of Minas Gerais*
Belo Horizonte, Brazil
figueiredo@dcc.ufmg.br

Larissa Rocha
*State University of Bahia*
Alagoinhas, Brazil
larissabastos@uneb.br

Sarah Nadi
*New York University Abu Dhabi*
Abu Dhabi, United Arab Emirates
sarah.nadi@nyu.edu

Fischer Ferreira
*Federal University of Itajubá*
Itajubá, Brazil
fischer.ferreira@unifei.edu.br

Geanderson Esteves
*Federal University of Minas Gerais*
Belo Horizonte, Brazil
geandersonesteves@gmail.com

*Abstract*—Large Language Models (LLMs) have gained attention for addressing coding problems, but their effectiveness in fixing code maintainability remains unclear. This study evaluates LLMs capability to resolve 127 maintainability issues from 10 GitHub repositories. We use zero-shot prompting for Copilot Chat and Llama 3.1, and few-shot prompting with Llama only. The LLM-generated solutions are assessed for compilation errors, test failures, and new maintainability problems. Llama with few-shot prompting successfully fixed 44.9% of the methods, while Copilot Chat and Llama zero-shot fixed 32.29% and 30%, respectively. However, most solutions introduced errors or new maintainability issues. We also conducted a human study with 45 participants to evaluate the readability of 51 LLM-generated solutions. The human study showed that 68.63% of participants observed improved readability. Overall, while LLMs show potential for fixing maintainability issues, their introduction of errors highlights their current limitations.

## I. INTRODUCTION

Code maintainability is important because it affects how easily the code can be understood, changed, and improved [1]. Poor maintainability can increase development costs, reduce software quality [2], and lead to slower delivery of new features [3]. Addressing maintainability issues involves refactoring the code to improve its structure, readability, and adherence to best practices. However, fixing maintainability issues can be challenging, because a specialist review is expensive and slow, while automated tools are imprecise and require human interpretation [3]. Code smells are a common type of maintainability issue, with examples including methods that become complex and take on too many responsibilities, code that is no longer used, or instances where the same code snippets are repeated twice or more [4, 5, 6].

Recent advances in large language models (LLMs) have generated much interest in their use for coding problems [7, 8, 9, 10, 11]. These models have shown remarkable capabilities in generating code, repairing bugs, and conducting software tests, but using LLMs to refactor maintainability issues is under-explored and lacks relevant data [12]. In contrast to traditional automated tools that adhere to rigid rules, LLMs can provide an innovative approach to addressing maintainability issues. Their ability to understand complex contexts allows them to generate flexible and adaptable solutions.

Furthermore, LLMs early investigations typically employ controlled scenarios to measure LLM capabilities [13, 14, 15, 16]. Despite these initial results, it is essential to understand the effectiveness of LLMs in addressing issues within real-world software projects [17, 18]. This type of software projects introduces numerous challenges for LLMs, including the need to understand and navigate codebases, adhere to various coding standards, and maintain compatibility with existing systems. Therefore, it is crucial to investigate how LLMs can resolve issues without introducing new errors or unintended behavior [8, 10, 14, 19].

In this paper, we evaluate the effectiveness of using LLMs to fix maintainability issues within Java methods. Our goal is to understand which maintainability issues LLMs can fix and where they fail. We aim to provide a comprehensive assessment of the utility of LLMs in maintenance tasks.

To conduct our empirical study, we use SonarQube 10.3.0 [1] to collect 127 instances of maintainability issues out of 10 GitHub Java projects, which have recent development activity, strong community, and a test suite available. The instances of the detected issues correspond to violations of 10 unique SonarQube rules. Then, we experiment with a proprietary LLM, Copilot Chat (version 0.15.2)[2], and an open-source LLM, Llama 3.1 70B Instruct[3], to fix the issues. We employ a zero-shot approach for both LLMs and a few-shot prompting approach for Llama to evaluate their performance across different prompt configurations. We also conduct a human study involving 45 participants to evaluate the readability of the LLM-generated successful solutions. By involving human participants, we aim to capture qualitative insights that complement the quantitative results of the LLMs performance.

---
[1]https://docs.sonarsource.com/sonarqube/latest/
[2]https://tinyurl.com/vud9rnsf
[3]https://ai.meta.com/blog/meta-llama-3-1/

Our results show that Llama with few-shot prompting fixed more methods than Copilot Chat and Llama with zero-shot. However, these configurations caused compilation errors and test failures in several instances, and failed to fix some methods. Furthermore, Copilot Chat and Llama zero-shot introduced new maintainability issues. Human evaluation showed that most LLM-generated solutions were more readable than the original code with maintainability issues.

Our study highlights the potential and limitations of LLMs in fixing real-world software maintenance issues. Copilot Chat and Llama showed limited effectiveness in fixing maintainability issues in real-world projects, with a tendency to introduce new problems. This underscores the need for continued human oversight and further refinement of LLM capabilities.

In summary, this paper makes the following contributions:

1) Quantitative and qualitative analysis of LLM-generated solutions comparing different prompting approaches (zero-shot and few-shot) to fix maintainability issues.
2) Insights on the limitations of LLMs in fixing maintainability issues, with code examples for common errors (see IV-A and replication package).
3) A public dataset consisting of pairs of methods with maintainability issues and their LLM-generated solutions, created using three approaches, available for replication and use in other experiments.

**Replication Package.** The online artifacts used in this study are available at https://zenodo.org/records/13921292.

## II. RELATED WORK

*Correctness Evaluation of Copilot.* Several previous empirical studies [8, 10, 20, 21] have aimed to evaluate Copilot's performance in software development tasks. For instance, Nguyen and Nadi [5] empirically evaluated the correctness and understandability of Copilot's suggested code in four different languages (Python, Java, JavaScript, and C) for 33 LeetCode questions. Their results showed that Java suggestions had the highest likelihood of being correct and that Copilot's suggestions have low cyclomatic and cognitive complexity (median 5 and 6, respectively). To repay self-admitted technical debts (SATD), O'Brien et al. [8] relied on Copilot to automatically generate 1,140 code bodies for TODO comments. From a different perspective, Pearce et al. [10] investigated the security of Copilot's code contributions and the conditions that cause GitHub Copilot to recommend insecure code. In their context, they found that up to 40% of Copilot-generated code could be vulnerable. Unlike these published studies [10, 21], our paper focuses on whether and how Copilot solves maintainability issues in actual software projects from GitHub.

*Correctness Evaluation of Llama.* Several studies have compared the effectiveness of Llama with proprietary LLMs [22, 23]. Jensen et al. [22] evaluated the effectiveness of proprietary and open-source LLMs for identifying code with security vulnerabilities. The study used zero-shot prompting to assess Llama 2's effectiveness in providing detailed descriptions of security vulnerabilities. Zhu et al. [23] assessed the capabilities of Code Llama and other LLMs for code summarization. In most cases, GPT-3.5 and GPT-4 produced better results for Java, but sometimes CodeLlama outperforms other LLMs for Python. When comparing different prompting techniques, CodeLlama, when using few-shot prompting, had better results compared to zero-shot learning in most cases. Like these studies, our work compares proprietary and open-source LLMS, including different prompting approaches, but in the software maintainability context.

*Evaluation of LLMs for Code Refactoring.* Initial studies evaluated the effectiveness of LLMs for code refactoring [18, 24, 25]. Choi et al. [18] propose an iterative project-level code refactoring process to reduce complexity by identifying the methods with the highest Cyclomatic Complexity and performing refactoring on them using ChatGPT 3.5. The results show that the average Cyclomatic Complexity is reduced with several iterations. Pomian et al. [24] propose a tool named EM-Assist that utilizes ChatGPT 3.5 to generate refactoring suggestions for methods that require the Extract Method refactoring technique. The tool ranks the solutions to provide developers with high-quality options. The results demonstrate that EM-Assist achieves a recall rate of 53% to fix complex methods. Shirafuji et al. [25] propose a method to select the best-suited code refactoring examples used for few-shot prompts to reduce Cyclomatic Complexity, using ChatGPT 3.5. The results show that their method can reduce the complexity. These studies are very focused on Cyclomatic Complexity issues and Extract Method techniques. They also focus on ChatGPT 3.5. Our study evaluates different types of maintainability issues, refactoring techniques, and LLMs.

## III. STUDY DESIGN

In this section, we present the design of our study, which aims to evaluate the effectiveness of LLMs in addressing maintainability issues. To identify these issues, we utilize SonarQube, a tool that detects violations of predefined *rules*, such as *"String literals should not be duplicated"*. When a rule is violated, SonarQube reports it as an issue. Maintainability issues are categorized as a type of code smell by SonarQube [26]. Our evaluation focuses on assessing the capability of LLMs to fix different types of maintainability issues by addressing the following research questions:

**RQ1. To what extent can LLMs fix maintainability issues?** For each Java method that has maintainability issues detected by SonarQube, we assess how many of these the LLMs can fix. We assess not only whether the LLM fixed the original issue, but also if the code passed the build process and if no new maintainability issues were introduced.

**RQ2. What are the main errors made by LLMs when attempting to fix maintainability issues?** We evaluate the behavior of LLMs in addressing different types of maintainability issues, highlighting the main refactoring techniques used and the most common errors. For each maintainability category, we provide a code example illustrating the most common mistake.

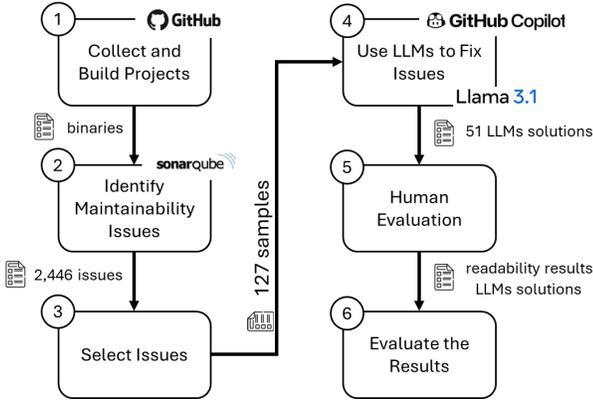

Fig. 1. Steps to evaluate if LLM can automatically fix maintainability issues.

TABLE I
PROJECTS WITH STUDY ISSUES STATISTICS AND POPULARITY

| Projects and Versions | #Issues | Samples | LOC | Stars | Contributors |
|---|---|---|---|---|---|
| Apollo 2.2.0 | 119 | 6 | 52k | 29.1k | 154 |
| Byte Buddy 1.14.13 | 317 | 11 | 184k | 6.2k | 98 |
| Google Java Format 1.21.0 | 101 | 11 | 16k | 5.6k | 93 |
| Google Jimfs 1.3.0 | 47 | 6 | 17k | 2.4k | 27 |
| Google Guava 33.0.0 | 782 | 16 | 97k | 50.1k | 308 |
| Google Guice 7.0.1 | 173 | 14 | 27k | 12.5k | 77 |
| Jitwatch 1.4.9 | 365 | 24 | 84k | 3.1k | 32 |
| Jsoup 1.18.1 | 171 | 19 | 32k | 10.9k | 108 |
| Zxing 3.5.4 | 264 | 16 | 39k | 32.7k | 126 |
| Webmagic 0.10.1 | 107 | 4 | 16k | 11.4k | 55 |
| total | 2,446 | 127 | | | |

**RQ3. To what extent do developers find LLM solutions more readable?** We assess the opinions of 45 participants by comparing 51 pairs of methods. Each pair has methods with maintainability issues and their LLM-generated solution.

Figure 1 shows the study steps, discussed in detail below in the following subsections.

### A. Step 1: Collect and Build Projects

Our goal in this study is to analyze how the LLMs fix maintainability issues in real-world Java projects. To facilitate our evaluation, these projects must have automated build pipelines, such that we can automatically verify the LLM-generated solutions. Accordingly, we randomly selected 10 open-source Java software projects hosted on GitHub that meet the following criteria: (1) at least three quarters (75%) of its code is in Java; (2) the project is actively maintained (for this, we consider projects with updates from 2023 onwards); (3) the project has broad recognition in the community (we select software projects with at least 1,000 stars in GitHub); (4) the project should have tests. We chose Java due to its widespread use in both academia and industry [27, 28, 29], focusing on Maven, a popular build system [30], as its logs help identify compilation errors and test failures for our analysis.

### B. Step 2: Identify Maintainability Issues

We use SonarQube to identify maintainability issues, as it is one of the most widely used static analysis tools among GitHub users and in the industry [31, 32, 33]. To avoid selecting issues specific to the Java language, we focus on evaluating general maintainability rules that are common across languages, enabling the study to be replicated in other programming languages. We identify SonarQube rules shared by the three most-used languages on GitHub (JavaScript, Python, and Java) [29], resulting in a total of 79 rules. We run SonarQube on each of the 10 selected projects and collect all violations[4] of these 79 rules, focusing only on the main source code and excluding test code. Each reported issue contains (1) the violated rule; (2) the file path with the issue and; (3) the line in which the issue occurs.

[4]In this paper, *rule violation* and *issue* are used interchangeably.

Table I presents information on projects and the number of issues evaluated in the study. The first column contains the name of the repository and the version we analyzed. The second column is the total number of maintainability issues in the main source code when considering the 79 selected rules. The third column is the number of issues we evaluated in the study per the selection criteria we discuss in Section III-C. The fourth column is the number of lines of code (disregarding comments) of the projects. The fifth column is the number of stars that the project had in 2024-10-11. The sixth column is the number of project contributors in 2024-10-11.

### C. Step 3: Select Issues

Table II shows the details of the 127 selected issues and the rules they violate. The first column contains the acronym for each rule, which we use throughout the paper. The second column contains the complete rule description and the third column shows the number of samples per rule. The last column shows the number of unique projects violating the rule.

We collected 2,446 maintainability issues detected in the main source code of the 10 projects. For feasibility, we define criteria to obtain representative samples: (1) randomness: for each rule, we choose the issues randomly; (2) sample size: as this is an exploratory study aiming to identify general patterns, we consider a sample size with a confidence level of 90%, allowing for a 7% margin of error; (3) minimum filtering: for each type of rule, we consider only cases that have at least 5 issues; (4) variability: each rule violation is represented in at least 3 projects. Using these criteria, we are left with a representative sample of 127 maintainability issues that represent violations of 10 most frequent rules.

### D. Step 4: Using an LLM to Fix Issues

We use two LLMs to address maintainability issues: GitHub Copilot Chat and Meta Llama 3.1 70B Instruct. To evaluate the effectiveness of LLMs, we used different prompting approaches with zero-shot and few-shot learning. Zero-shot learning occurs when the prompt is provided without any examples, while few-shot learning presents a limited number of examples within the prompt [34]. The approaches are: (1) Copilot Chat with zero-shot prompting; (2) Llama 3.1 70B Instruct with zero-shot prompting; (3) Llama 3.1 70B Instruct with few-shot prompting. For brevity, we refer to these approaches from now on as *Copilot Chat*, *Llama zero-shot*, and *Llama few-shot*, respectively. Copilot Chat and Llama have significantly different interfaces, so we distinctly experiment with each.

TABLE II
DESCRIPTION OF THE 127 SELECTED MAINTAINABILITY ISSUES

| Acronyms | Rules | Issues | Projects Coverage |
|---|---|---|---|
| CCM | Cognitive Complexity of methods should not be too high | 26 | 9 |
| GET | Generic exceptions should never be thrown | 13 | 6 |
| MIS | Mergeable if statements should be combined | 9 | 5 |
| CCO | Sections of code should not be commented out | 13 | 6 |
| SLD | String literals should not be duplicated | 19 | 7 |
| TON | Ternary operators should not be nested | 8 | 5 |
| TUT | Track uses of "TODO" tags | 16 | 6 |
| TBS | Two branches in a conditional structure should not have exactly the same implementation | 5 | 3 |
| UAR | Unused assignments should be removed | 8 | 3 |
| UMP | Unused method parameters should be removed | 10 | 5 |
| | total | 127 | |

**Copilot Chat.** Figure 2 shows the Copilot Chat dialog and an example of the prompt that we use in the study. We highlight the entire method by which SonarQube reported the issue; this defines the *context* for Copilot Chat. We then insert a request in the Copilot Chat dialog box using natural language: *"In this class, the method **method_name** has the following issue **rule_description**. Can you identify and /fix it?"*. The method_name and rule_description are variables. The term /fix is a slash command that informs Copilot Chat about our intention of fixing the provided code. Most of the time, in addition to the code suggestion, the Copilot Chat explains what will be done. We register this explanation for analysis.

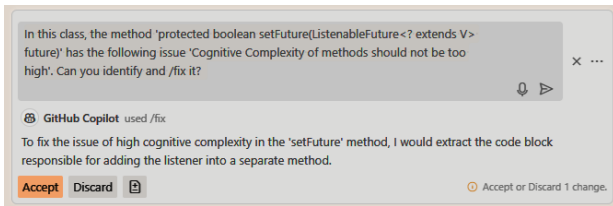

Fig. 2. Copilot Chat with a prompt example and solution explanation.

**Llama 3.1-70B-Instruct.** To use Llama 3.1 70B Instruct, we use HuggingChat[5] because the model requires more than 140GB of VRAM. Unlike ChatGPT[6], a copied code must be manually pasted into the project. We use the same prompt that we use in Copilot Chat in Llama, but we also have to include the snippet to be refactored in the prompt. The difference between zero-shot and few-shot learning is that, in few-shot learning, we provide three examples (inspired by [25, 35]) of code pairs, each consisting of the code with the issue and its corresponding refactored version, within the prompt. The few-shot examples were extracted from SonarQube documentation and community contributions, and are included in the replication package.

In general, proprietary LLMs like Copilot Chat outperform open-source models [22, 23]. We use zero-shot on Copilot to evaluate its effectiveness without examples. For Llama, we apply zero-shot to check if its performance would be inferior or comparable to Copilot, and then use few-shot to see if Llama could surpass it. This approach allowed for a direct comparison between the two models and highlighted the potential of open-source LLMs. We accept whatever suggestion was returned by the LLMs, and then compile the project and run the tests. If the project with the change compiles and tests successfully, we submit it for a new analysis by SonarQube and record whether the issue was fixed (disappeared) or if any new issues appear, which we refer to as *code degradation*. SonarQube outputs were evaluated by the authors, with doubts discussed in meetings. The information of the original source code, solution by LLM, compilation and test logs, and SonarQube new analysis is available in the replication package.

Before the main experiment, we conducted a pilot study analyzing the CK Metrics repository [7] to evaluate if our study design is feasible. Without further obstacles, we proceeded to define the design of the human evaluation.

### E. Step 5: Human Evaluation

Our goal is to have real developers evaluate the readability of the LLM-generated solutions. We invited 45 participants, comprising undergraduate and graduate (MSc and PhD) students from the Computer Science department at our university, to take part in this study, which was conducted as part of a course project. To ensure adequate feedback on each solution, we ensured that every solution was assessed by at least two participants. For this user study, we considered only the LLM solutions that successfully fixed the maintainability issues. Accordingly, we randomly selected 51 samples out of Copilot Chat, Llama zero-shot, and Llama few-shot to represent 9 categories of SonarQube rules. The only rule that was not considered in this step was UMP because we obtained only one successful LLM-generated solution from a single strategy.

For each selected method, we created a survey containing: (1) a *method pair* with the original method (with maintainability issues) and the LLM-modified version; (2) instructions for the participant to assume the role of a GitHub repository maintainer and choose the more readable method [36]; and

---
[5] https://huggingface.co/chat/
[6] https://chatgpt.com/
[7] https://github.com/mauricioaniche/ck

(3) a list of questions asking participants to (a) identify the differences between the methods, (b) indicate which method is more readable, and (c) justify their choice. Participants could also consider the methods equivalent or state they could not evaluate the differences between them. The order of the original and LLM-generated codes presented was randomized. After designing the survey[8], we conducted a pilot study with two MSc students to assess whether the questions were clear. Based on their feedback, we adjusted the survey text.

When conducting the study, we first briefly explained the provided instructions of the study and read the survey questions to clarify any questions participants might have. We collected for each method pair, the number of votes for each method (original and LLM codes), cases when the methods were judged as equivalent, and when the participant was not able to give an opinion.

## IV. EVALUATION

### A. Effectiveness of an LLM in fixing maintainability issues

Figure 3 shows the results for the three LLM configurations. The X-axis represents the status of the LLM solution: (1) *fixed*: methods that were fixed by the LLMs without introducing any errors or failures; (2) *not fixed*: methods that were neither fixed nor introduced errors or failures; (3) *compilation error*: methods that caused compilation errors after the LLMs changes; (4) *test failure*: methods that caused test failures after the LLMs changes; (5) *degraded*: methods where new maintainability issues were introduced after the LLMs changes (6) *no suggestion*: methods for which the LLMs did not suggest any fix.

From the 127 samples in the study, the best effectiveness in fixing maintainability issues was achieved by the Llama few-shot approach, which fixed 57 (44.9%) methods, followed by Copilot Chat with 41 (32.29%) methods, and Llama zero-shot with 38 (30%) methods. The worst result of *not fixed* comes from Copilot Chat with 23 (18.11%) methods, followed by Llama zero-shot with 22 (17.22%) methods and Llama few-shot with 12 (9.44%) methods. Concerning the cases with compilation errors, the worst case is the Llama zero-shot approach with 42 (33%) methods, followed by Llama few-shot with 37 (29.13%) methods, and Copilot Chat with 32 (25.2%) methods. Regarding LLM-generated solutions with test failures, Llama zero-shot has 19 (15%) methods, Llama few-shot has 18 (14.17%) methods, and Copilot Chat has 9 (7.08%) methods. Only Copilot Chat and Llama zero-shot have degraded methods, 6 (4.72%) and 5 (4%) respectively. Finally, in cases where LLMs did not suggest any solution, Copilot Chat has 16 (12.6%) methods, Llama zero-shot (0.78%) has 1 method, and Llama few-shot has 3 (2.36%) methods.

[8]https://forms.gle/BMgzNWBcQV32HYK77

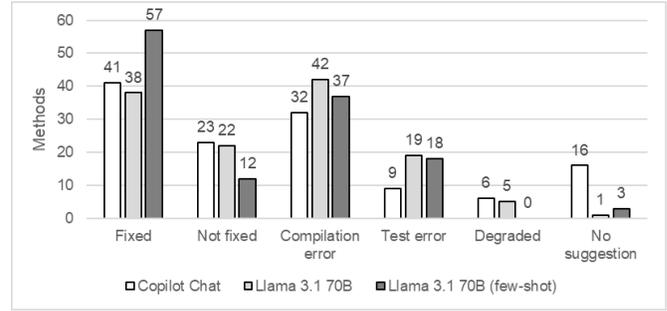

Fig. 3. The result of fixing 127 methods with 3 LLMs aprroaches

**Answer of RQ1:** Our evaluation reveals that Llama few-shot demonstrated the highest effectiveness in addressing maintainability issues, successfully fixing 57 out of 127 methods (44.9%). In comparison, Copilot Chat and Llama zero-shot fixed 41 (32.29%) and 38 (30%) methods, respectively. Notably, in addition to the Llama few-shot yielding the highest fixing rate, it also introduces fewer new maintainability issues compared to the other methods, indicating its superior performance in both fixing existing problems and maintaining code quality.

### B. Refactoring Techniques and common LLM errors in maintainability fixes

Table III shows the results categorized by the 10 SonarQube rules. The first column shows the status of the LLM solution. The second column lists the LLMs and the approaches used (zero-shot or few-shot). The remaining columns display the acronyms of the rules, as shown in Table II. Each numeric value represents the number of methods. We highlight some results in each status in bold. We want to show a **higher** number in the fixed status and a **lower** number in the other statuses.

We analyzed 127 LLM-generated solutions out of the 3 strategies to evaluate the effectiveness of LLMs and to understand what types of refactoring techniques they use.

Hallucination in LLMs occurs when the model generates incorrect or irrelevant information, creating responses that appear plausible but are not grounded in real data [37]. This concept is crucial for this section because we provide some code examples to exemplify common errors when LLMs fail to fix maintainability issues. Due to space limitations in the paper, we use [...] to omit parts of the code that are not important for the discussion. We present the snippet of the original code and the refactored code immediately after the comments '// original' and '// LLM-generated solution', respectively. We now discuss our results.

**The cognitive complexity of methods should not be too high (CCM).** This issue occurs when the method control flow is hard to understand. We evaluated 26 methods that violated the CCM rule. Copilot Chat shows the best effectiveness, fixing 7 methods (26.9%), followed by Llama, which fixes 6 methods (23%) in both zero-shot and few-shot approaches. The effectiveness rate is low. Only 2 (7.7%) methods are fixed by all strategies, and at least one LLM fixed 10 (38.46%) methods. Regarding the refactoring strategies used to fix the issue, Copilot Chat uses the *Extract Method* [4] 6 times and *Split Conditional* [4] once, while Llama zero-shot and few-shot use the Extract Method every time. Concerning failed cases, the most common type is *compilation error*, with 15 occurrences in Llama few-shot and 11 in the other approaches. The most frequent *compilation errors* are: *Cannot find symbol*, *Incompatible types*, and *Element is already defined*. Code 1 shows an example where the *Extract Method* refactoring technique was applied to the *put* method, creating new methods, including *overwriteEntry*. Note that in the original method, the variable *prev* is declared as a *DirectoryEntry*, while in the LLM-generated solution, *prev* has become an attribute of the *DirectoryEntry* object. This example illustrates a common type of wrong logic interpretation caused by LLMs when attempting to fix CCM issues. Another important observation is the high number of LLM-generated solutions that change the method behavior for CCM: 7, 5, and 4 for Llama zero-shot, Llama few-shot, and Copilot zero-shot, respectively. These errors suggest that LLMs struggle to maintain the context and dependencies when moving code elements.

TABLE III
RESULTS OF FIXING 127 METHODS: COMPARISON OF ALL LLM APPROACHES BY RULE VIOLATION

| Status | LLM | CCM | GET | MIS | CCO | SLD | TON | TUT | TBS | UAR | UMP |
|---|---|---|---|---|---|---|---|---|---|---|---|
| Fixed | Copilot Chat | 7 | 3 | 4 | 9 | 3 | 6 | 4 | 2 | 3 | 0 |
| | Llama 3.1 70B (zero-shot) | 6 | 4 | 5 | 6 | 5 | 5 | 2 | 3 | 1 | 1 |
| | Llama 3.1 70B (few-shot) | 6 | 4 | 6 | 12 | 6 | 6 | 12 | 2 | 3 | 0 |
| Not fixed | Copilot Chat | 2 | 3 | 1 | 0 | 13 | 0 | 2 | 1 | 0 | 1 |
| | Llama 3.1 70B (zero-shot) | 1 | 3 | 2 | 4 | 5 | 0 | 1 | 1 | 4 | 1 |
| | Llama 3.1 70B (few-shot) | 0 | 1 | 1 | 1 | 6 | 0 | 0 | 1 | 2 | 0 |
| Compilation error | Copilot Chat | 11 | 3 | 3 | 1 | 0 | 2 | 3 | 0 | 1 | 8 |
| | Llama 3.1 70B (zero-shot) | 11 | 4 | 1 | 3 | 4 | 2 | 9 | 0 | 1 | 7 |
| | Llama 3.1 70B (few-shot) | 15 | 7 | 1 | 0 | 2 | 1 | 0 | 0 | 2 | 9 |
| Test failure | Copilot Chat | 4 | 0 | 1 | 0 | 0 | 0 | 0 | 1 | 3 | 0 |
| | Llama 3.1 70B (zero-shot) | 7 | 1 | 1 | 0 | 3 | 1 | 4 | 1 | 0 | 1 |
| | Llama 3.1 70B (few-shot) | 5 | 1 | 1 | 0 | 4 | 1 | 3 | 2 | 1 | 0 |
| Degraded | Copilot Chat | 1 | 1 | 0 | 0 | 2 | 0 | 2 | 0 | 0 | 0 |
| | Llama 3.1 70B (zero-shot) | 1 | 1 | 0 | 0 | 1 | 0 | 0 | 0 | 2 | 0 |
| | Llama 3.1 70B (few-shot) | 0 | 0 | 0 | 0 | 0 | 0 | 0 | 0 | 0 | 0 |
| No suggestion | Copilot Chat | 1 | 3 | 0 | 3 | 1 | 0 | 5 | 1 | 1 | 1 |
| | Llama 3.1 70B (zero-shot) | 0 | 0 | 0 | 0 | 1 | 0 | 0 | 0 | 0 | 0 |
| | Llama 3.1 70B (few-shot) | 0 | 0 | 0 | 0 | 1 | 0 | 1 | 0 | 0 | 1 |

**Generic exceptions should never be thrown (GET).** This issue occurs when the method throws generic exceptions instead of specific ones. We evaluated 13 methods that violated the GET rule. Llama has the best effectiveness for zero-shot and few-shot approaches, both fixed 4 (30.7%) methods, while Copilot Chat fixed 3 (23%) methods. Like in CCM, the effectiveness rate is still low. Only 2 (15.3%) methods are fixed by all strategies, and at least one LLM fixes 6 (46.1%) methods. All method fixes use the *Change Thrown Exception Type* [38] refactoring technique, with one exception that uses *Add Thrown Exception Type* [39] by creating two exception classes.

```
// Original:
private void put(DirectoryEntry entry, boolean
   overwriteExisting) {
    DirectoryEntry prev = null;
    [...]
    prev = curr;
    [...]
    if (prev != null) {
        prev.next = entry;
    }
    [...]
}

// LLM-generated solution:
private void put(DirectoryEntry entry, boolean
   overwriteExisting) {
    [...]
    overwriteEntry(existingEntry, entry);
    [...]
}

private void overwriteEntry(DirectoryEntry existingEntry,
   DirectoryEntry newEntry) {
    [...]
    if (existingEntry.prev!= null) {
        existingEntry.prev.next = newEntry;
    }
    [...]
}
```

Code 1: CCM example of hallucination.

Cases that use *Change Thrown Exception Type* change *RuntimeException* for *IllegalStateException*. Most failed cases are caused by compilation errors, specifically when LLM changes the signatures with the throws command (most common in Llama few-shot) and when using exceptions without importing their classes. Llama few-shot, Llama zero-shot, and Copilot Chat have 7 (53.8%), 4 (30.7%) and 3 (23%) LLM-generated solutions classified as *compilation error*, respectively. To use LLMs to fix GET issues, developers should specify their exceptions, indicate what type of refactoring they want to use

```
1  // Original:
2  if ((sawNewlines && isComment(text)) ||
   ↪    wanted.wanted().orElse(sawNewlines)) {
3      ++newlinesPending;
4  }
5
6  // LLM-generated solution:
7  if (sawNewlines || isComment(text) ||
   ↪    wanted.wanted().orElse(false)) {
8      ++newlinesPending;
9  }
```

Code 2: MIS example of hallucination.

```
1  //Original:
2  // We extend the recursive expression for the one-variable
   ↪   case at Art of Computer Programming
3  // vol. 2, Knuth, 4.2.2, (16) to the two-variable case. We
   ↪   have two value series x_i and y_i.
4  // We define the arithmetic means X_n = 1/n \sum_{i=1}^n
   ↪   x_i, and Y_n = 1/n \sum_{i=1}^n y_i.
5
6  //LLM-generated solution:
7  public double previousMean() {
8      if (count() > 1) {
9          return (sum() - getLast()) / (count() - 1);
10     } else {
11         return 0; // or throw an exception, depending on
   ↪           your requirements
12     }
13 }
```

Code 3: CCO example of hallucination.

(whether to add or change the exception type), instruct the prompt not to change method signatures, and be careful about importing new classes.

**Mergeable if statements should be combined (MIS).** This issue occurs when a method has excessive nested conditionals, increasing the code complexity and merge conflict resolution [40]. We evaluated 9 methods that violated the MIS rule. The effectiveness of the LLMs in fixing these cases is as follows: 6 methods (66.7%) for Llama few-shot, 5 methods (55.6%) for Llama zero-shot, and 4 methods (44.4%) for Copilot Chat. Only 4 methods (44.4%) were fixed by all strategies, while at least one LLM successfully fixed 7 methods (77.8%). All successful solutions generated by the LLMs use the *Consolidate Conditional Expression* refactoring [4]. Compilation errors and test failures occur when LLMs hallucinate. Code 2 shows an example of Llama zero-shot hallucination, because it changes the fallback logic from using *sawNewlines* to always using false, altering the original behavior, causing *test failure*. The best way to fix MIS using LLMs is through an iterative approach [18] combined with automated tests to avoid compilation errors and ensure the correct method behavior.

**Sections of code should not be commented out (CCO).** This issue occurs when the method contains commented-out code, which makes it difficult to read. We evaluated 13 methods that violate the CCO rule. Llama few-shot and Copilot Chat demonstrate good effectiveness in addressing this issue, fixing 12 methods (92.3%) and 9 methods (69%), respectively. Both prioritize using *Removing Comment* refactoring to fix the problem. On the other hand, Llama zero-shot sometimes uncomments the code or attempts to implement it. As a result, it fixed only 6 methods (46.1%). Code 3 shows an example where Llama zero-shot implements the commented code. In this case, the LLM-generated solution included a new method named *previousMean*, called within the method, changing the code logic. These errors do not occur in Llama few-shot, as the examples in the few-shot explicitly instruct the deletion of the comment to fix the CCO issue. To fix CCO using LLMs, a few-shot with examples of comments removal, followed by instructions not to implement commented-out code, is enough. Our study, along with other studies [8], show that LLMs are still not successful at implementing code from comments.

**String literals should not be duplicated (SLD).** This issue occurs when a constant is used as a string multiple times in a method. We evaluated 19 methods that violate the SLD rule. The most common refactoring technique to fix SLD used by LLMs is *Extract Constant* [41]. The number of *fixed* methods for the Llama zero-shot and few-shot scenarios, 5 and 6 respectively, is equal to the number of methods *not fixed*. The results are worse for Copilot Chat, with 3 *fixed* methods and 13 *not fixed*. The identification of the violation of SLD by SonarQube generates some disagreements about false positives within the SonarQube community[9]. This is the only rule with at least one LLM response labeled as *no suggestion*. A Llama output reports *"the issue is not in the provided code snippet"*. This is the behavior we expect for false-positive cases, but instead, LLMs often generate uncommon suggestions, even for these cases, causing hallucinations. For example, Code 4 shows strings in the original code labeled as SLD issue by SonarQube, but each word appears only once in the method. The LLM output reports that *"I extract the repeated string literal into a constant DUPLICATE_FIELD_MESSAGE"*, but we notice that this constant does not exist in the LLM-generated solution. Instead, it creates a *message* variable, and upon secondary analysis, the SLD issue remains. Table III provides the values of all results. From this and other similar examples, and the number of *not fixed* methods, we assume that LLMs interpret the SLD issue differently than SonarQube. An approach to be tested would be to apply prompt engineering that specifies the need to count the number of repeated strings and request the creation of global variables only for those that exceed a developer-defined limit.

**Ternary operators should not be nested (TON).** This issue occurs when ternary operators are hard to read and can make the order of operations complex to understand. We evaluate 8 methods that violate the TON rule. Copilot Chat and Llama few-shot *fixed* 6 (75%) methods and Llama zero-shot *fixed* 5 (62.5%) methods. The LLM-generated solutions use *Decompose Conditional* [4] for refactoring TON issues. Code 5 shows an example of *Decompose Conditional* used by Llama few-shot. Compilation and test errors are caused by the difficulty in interpreting logic and rewriting code by the LLMs. Like for CCM and MIS, an iterative approach, followed by automated tests, are imporant to use LLMs for TON refactoring.

---

[9]https://community.sonarsource.com/t/java-duplicated-string-literals/46443

```
1  // Original:
2  else {
3      throw new IllegalStateException("Field " +
       ↪  token.getName()
4              + " for " + this
5              + " already mapped to " + previous
6              + " and not " + value);
7  }
8
9  // LLM-generated solution:
10 else {
11     String fieldName = token.getName();
12     String message = String.format("Field '%s' for %s
       ↪  already mapped to %s and not %s", fieldName, this,
       ↪  previous, value);
13     throw new IllegalStateException(message);
14 }
```

Code 4: SLD example of hallucination.

```
1  // Original:
2  size += (characterLength / 2) * 11;
3  size += (characterLength % 2) == 1 ? 6 : 0;
4
5  // LLM-generated solution:
6  size += (characterLength / 2) * 11;
7  if ((characterLength % 2) == 1) {
8    size += 6;
9  }
```

Code 5: TON example of LLM-generated solution.

```
1  // Original:
2  else {
3      // TODO(cushon): don't break after here for consistency
       ↪  with for, while, etc.
4      builder.breakToFill();
5      builder.open(ZERO);
6      scan(ex, null);
7      builder.close();
8  }
9
10 // LLM-generated solution:
11 else {
12     // TODO(cushon): don't break after here for consistency
       ↪  with for, while, etc.
13     builder.breakToFill();
14     builder.open(ZERO);
15     scan(ex, null);
16     builder.close();
17 }
18
19 [...]
20
21 // Track uses of TODO tags
22 if (node.getBlock().toString().contains("TODO")) {
23     // Perform tracking actions here
24 }
```

Code 6: TUT example of hallucination.

```
1  // Original:
2  else if (t.isEndTag() &&
   ↪  (inSorted(t.asEndTag().normalName(),
   ↪  BeforeHtmlToHead))) {
3      return anythingElse(t, tb);
4  } else if (t.isEndTag()) {
5      tb.error(this);
6      return false;
7  } else {
8      return anythingElse(t, tb);
9  }
10
11 // LLM-generated solution:
12 else if (t.isEndTag()) {
13     tb.error(this);
14     return false;
15 } else {
16     return anythingElse(t, tb);
17 }
```

Code 7: TBS example of hallucination.

**Track uses of TODO tags (TUT).** This issue occurs when the method contains TODO commented-out code, which makes it difficult to read. We evaluate 16 methods that violate the TUT rule. The three approaches have different behaviors for this type of issue. Copilot Chat fixed 4 (25%) methods, did not suggest any solution for 5 (31.2%) methods, and degraded the code when tried to fix 2 (12.5%) methods. Code 6 shows an example of code degradation where Copilot Chat did not fix the TUT issue and created an empty conditional, adding a new issue. Llama zero-shot tried to implement the TODO comment in most cases because that caused 9 (56.2%) compilation errors, 4 (25%) test failures, and only 2 (12.5%) fixed methods. The examples of few-shot adopt delete the comment as TUT solution, then Llama few-shot has the best effectiveness with 12 (75%) fixed methods. As with CCO, a few-shot approach with examples of comment removal, followed by instructions not to implement commented-out code, could be a better option for fixing TUT issues.

**Two branches in a conditional structure should not have exactly the same implementation (TBS).** This issue occurs when the same code is duplicated in two or more separate branches of a conditional. We evaluate 5 methods that violate the TBS rule. Llama zero-shot fixed 3 methods and the others 2 methods. The most successful refactoring is *Consolidate Conditional Expression* [4]. The LLMs struggled to understand the conditional logic, resulting in some *not fixed* and *test failure* cases. In the original Code 7, *return anythingElse(t, tb);* repeat twice. Then, Llama few-shot removes the first conditional. But, the LLM removed a condition that checks if the EndTag is in BeforeHtmlToHead, redirecting the code to the else if (t.isEndTag()) block. This changes the intended logic and may lead to incorrect error handling. To refactor TBS using LLMs, it could be treated as a code clone problem. There are prompt-based approaches [42] and fine-tuning methods [43], but they need to be explored in the context of TBS.

**Unused assignments should be removed (UAR).** This issue occurs when the method contains dead stores, which are unnecessary assignments that reduce code clarity and waste resources [44] . We evaluate 8 methods that violate the UAR rule. Copilot Chat and Llama few-shot have the best effectiveness with 3 (37.5%) fixed methods, if compared with Llama zero-shot with 1 (12.5%) fixed method. The LLMs use *Remove Dead Code* [4] for UAR. These results are limited. LLMs have difficulties with deep logical reasoning and sometimes delete assignments without UAR issues. Code 8 shows an example where *afterFirstToken* attribute was deleted and the conditional was changed, causing a compilation error. Our advice is to use automated tests and iterative approaches, like for CCM, MIS, TON issues.

**Unused method parameters should be removed (UMP).** This issue occurs when the method contains unused method parameters (UMP), which do not contribute to the functionality and can hinder code readability. We evaluate 10 methods that

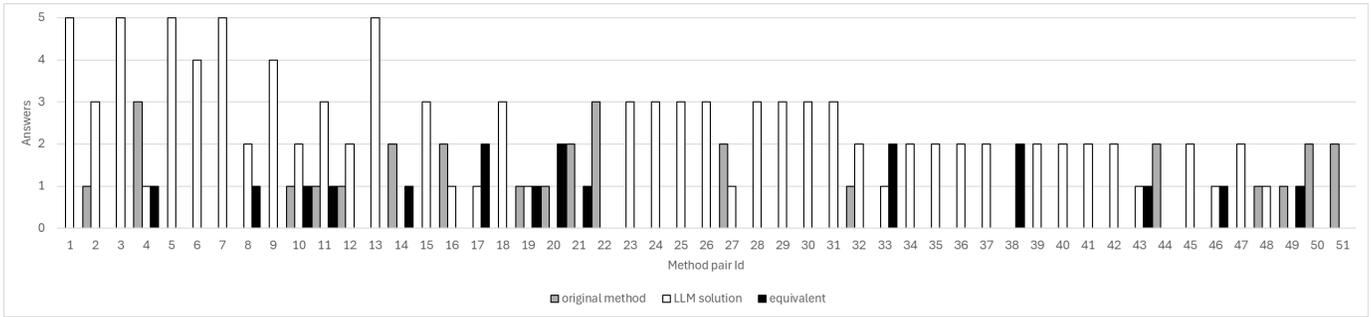

Fig. 4. Comparisons of answers by methods

```
1  // Original:
2  boolean afterFirstToken = false;
3  [...]
4  if (afterFirstToken) {
5    builder.breakOp(INDEPENDENT, " ", ZERO,
    ↪  Optional.of(breakBeforeType));
6  } else {
7    afterFirstToken = true;
8  }
9
10 // LLM-generated solution:
11 if (!typeAnnotations.isEmpty() ||
    ↪  !node.getTypeParameters().isEmpty()) {
12    builder.breakOp(INDEPENDENT, " ", ZERO,
    ↪  Optional.of(breakBeforeType));
13 }
```

Code 8: UAR example of hallucination.

```
1  // Original:
2  public ImmutableMap<String, ?> defaultValues(Map<String,
    ↪  ?> userDefaults) {
3      return ImmutableMap.of();
4  }
5
6  // LLM-generated solution:
7  public ImmutableMap<String,?> defaultValues(Map<String,?>
    ↪  userDefaults) {
8      return ImmutableMap.<String, Object>builder()
9          .putAll(userDefaults)
10         .build();
11 }
```

Code 9: UMP example of LLM-generated solution.

violate the UMP rule. The LLMs use *Remove Dead Code* [4] for UMP. The majority of LLM-generated solutions result in compilation errors because they change the signature of a method, causing errors in calls from other parts of the code. The error was already expected, as LLMs do not have an overview of the entire system. However, there was an exception where the UMP issue was fixed. Code 9 shows the LLM-generated solution where the method implements the use of a parameter that was not previously used. It is important to note that there were no failures in the tests, as the parameter may always receive blank values, but the behavior of both methods is not the same, even though the issue has been fixed. LLMs like Copilot Chat and Llama are not capable of acting at the project-level [45], and fixing UMP without this type of resource becomes unfeasible.

**Answer of RQ2:** LLMs often make errors in code refactoring, such as generating solutions that lead to compilation failures. They also change the method behavior, especially with complex methods. Furthermore, they introduce irrelevant constants or variables, which highlights their limitations in handling refactorings accurately.

*C. Readability for LLM solutions*

We evaluate the readability of the LLM's solutions by asking 45 participants to rate the readability of 51 solutions. We got 153 ratings in total. Among these 153 ratings: (1) 44 (28.76%) rated the original methods as more readable than the LLM's fixed version; (2) 91 (59.48%) rated the LLM's fixed version as more readable than the original method; (3) 18 (11.76%) considered both versions of the method equivalent; and (4) no one answer *I don't know*.

Figure 4 provides the detailed votes for each of the 51 comparisons. The X-axis represents the method pair ID, and the Y-axis represents the number of answers that voted for a given code, as described in the legend. Since we have at least two participants' rates for each solution, we also analyze the most voted choice for each compared pair (i.e., had the majority vote). Out of the 51 solutions, participants found the LLM fixed version of the code more readable in 35 cases (68.63%) while the original version of the code received the majority of votes in 7 cases (13.73%). Participants found no difference in readability between both versions of the code for 4 cases (7.84%) while 5 (9.80%) other cases were a draw.

Table IV shows how often an answer is the most chosen in a comparison (organized by rules). From this point of view, the *LLM solution* answer is the most readable for 8 rules, in the other 1 rule original method is the most readable. Regarding the readability performance of each LLM, Copilot Chat had 16 pairs of methods evaluated, with the LLM-generated solution being chosen 14 times, and the original method chosen twice. Llama (zero-shot) had 17 pairs of methods assessed, with the LLM-generated solution chosen 12 times, *equivalent* answered 3 times, and the original method chosen twice. Llama (few-shot) had 18 pairs of methods analyzed, with the LLM-generated solution preferred 9 times, a draw occurring 5 times, the original method chosen 3 times, and an equivalent preference between the methods once.

TABLE IV
NUMBER OF TIMES WHEN AN OPTION WINS A COMPARISON BY THE RULES

| rules | original method | LLM solution | equivalent | draw |
|---|---|---|---|---|
| CCM | 0 | 6 | 0 | 0 |
| GET | 1 | 3 | 0 | 1 |
| MIS | 1 | 5 | 0 | 0 |
| CCO | 2 | 3 | 0 | 1 |
| SLD | 0 | 4 | 1 | 1 |
| TON | 0 | 5 | 0 | 1 |
| TUT | 3 | 1 | 2 | 0 |
| TBS | 0 | 5 | 0 | 0 |
| UAR | 0 | 3 | 1 | 1 |
| TOTAL | 7 | 35 | 4 | 5 |

**Answer of RQ3:** Out of 51 pairs of methods evaluated by 45 developers, 35 (68.63%) considered the method with LLM solution more readable than the original method, while 7 (13.73%) found the original method more readable. Additionally, 5 (7.84%) deemed both methods equivalent, and in 5 (9.80%) cases, the comparison resulted in a draw. Copilot Chat has the more readable LLM-generated solutions, followed by Llama zero-shot and Llama few-shot.

## V. THREATS TO VALIDITY

**Internal Validity.** Internal validity concerns factors that could affect the results of our study without our knowledge. One primary concern is the use of two LLM tools, Copilot Chat and Llama 3.1 70B Instruct, for our analysis. Specific limitations or inherent biases in Copilot Chat and Llama 3.1 may influence our findings. Furthermore, the evaluation criteria and the method of assessing code maintainability might introduce subjective biases. Although we employed SonarQube for an objective measure of code quality and included multiple evaluators in the assessment process, differences in evaluators' experience and interpretation could still impact the results. Additionally, it is important to note that we lack visibility into the inner workings of Copilot Chat and Llama 3.1, adding an element of uncertainty to our analysis.

Additionally, the experimental setup, including the selection of maintainability issues and the criteria for success, may affect the outcomes. For instance, the decision to focus on specific SonarQube rules could introduce bias if these rules do not represent a comprehensive view of code maintainability. The potential for human error in interpreting and applying these rules during the evaluation process is another factor that could influence the internal validity of our study. Also, the prompt build for this study might introduce bias, as it attempted to emulate a developer's interaction with the tool, typically not employing more advanced techniques in its input.

**External Validity.** External validity addresses the extent to which our findings can be generalized beyond the specific context of our study. Our dataset comprised a selection of Java projects, which may limit the applicability of our results to other programming languages or types of software projects. To address this, we chose a diverse set of Java projects, aiming to cover different application domains and coding styles. The Java language also demonstrated good performance in studies using LLM [21]. However, future studies should include other programming languages and project types to enhance generalizability.

Moreover, the context in which the LLM was used (integrated within an IDE and prompted to fix specific issues) may not reflect other potential use cases of LLMs in software development. We attempted to replicate a realistic development environment but acknowledged that different settings and user interactions could lead to different outcomes. We selected Copilot Chat and Llama 3.1 due to their prominence in previous research and extensive training by OpenAI, Meta, and GitHub. These LLMs represent a state-of-the-art tool widely used in practice. To mitigate the impact of this choice, we carried out a study with human evaluations and maintained a critical perspective on Copilot and Llama performance throughout the study. Encouraging further studies in various environments and with different developer expertise levels will help validate our findings more broadly.

**Reliability.** Reliability concerns the consistency of our results. If the study were repeated under similar conditions, it should yield comparable results. The variability in Copilot Chat and Llama suggestions and the subjective nature of some evaluations could affect reliability. We mitigated this by standardizing the evaluation process, using a consistent dataset, and documenting our methodology. This ensures that future researchers can replicate our setup and obtain similar results. Additionally, the reproducibility of our results is dependent on the specific version of the tools and datasets used. To address this, we archived the versions of Copilot Chat, Llama, SonarQube, and the datasets used in our study, providing a reference for future studies. This approach helps maintain consistency even as tools and environments evolve.

## VI. CONCLUSIONS AND FUTURE WORK

This paper evaluated the effectiveness of LLMs in fixing maintainability issues by mining 127 methods with issues, corresponding to 10 SonarQube rules, from 10 GitHub repositories. Llama few-shot fixed 57 (44.9%) out of 127 methods, the Copilot Chat 41 (32.29%) and Llama zero-shot 38 (30%). Furthermore, in our human evaluation study, 68.6% of developers considered the LLM solutions more readable compared to the original methods. We conclude that although LLMs show potential in improving code readability and fixing maintainability issues, their effectiveness is limited, and they often introduce new errors or fail to fix maintainability issues. In our study, LLM performance in addressing maintainability issues fell short, but recent studies highlight fine-tuning as a promising optimization strategy [46, 47, 48, 49], which we aim to explore in our research. We also plan to explore other languages and different types of LLMs as part of these efforts.

**Acknowledgements.** This research was partially supported by Brazilian funding agencies: CNPq (Grant 312920/2021-0), CAPES, and FAPEMIG (Grant APQ-01488-24).

## REFERENCES

[1] M. Riaz, E. Mendes, and E. Tempero, "A systematic review of software maintainability prediction and metrics," in *3rd International Symposium on Empirical Software Engineering and Measurement (ESEM)*, 2009, pp. 367–377.

[2] A. Santana, E. Figueiredo, J. Pereira, and A. Garcia, "An exploratory evaluation of code smell agglomerations," *Software Quality Journal (SQJ)*, 2024.

[3] M. Schnappinger, A. Fietzke, and A. Pretschner, "Defining a software maintainability dataset: collecting, aggregating and analysing expert evaluations of software maintainability," in *IEEE International Conference on Software Maintenance and Evolution (ICSME)*, 2020, pp. 278–289.

[4] M. Fowler, *Refactoring: improving the design of existing code*. Addison-Wesley Professional, 2018.

[5] H. G. Nunes, A. Santana, E. Figueiredo, and H. Costa, "Tuning code smell prediction models: A replication study," in *Proceedings of the 32nd IEEE/ACM International Conference on Program Comprehension (ICPC)*, 2024.

[6] A. Santana, J. A. Pereira, and E. Figueiredo, "Impact of code smell agglomerations on code stability," in *Proceedings of the 40th International Conference on Software Maintenance and Evolution (ICSME)*, 2024.

[7] N. Al Madi, "How readable is model-generated code? examining readability and visual inspection of github copilot," in *Proceedings of the 37th IEEE/ACM International Conference on Automated Software Engineering (ASE)*, 2022, pp. 1–5.

[8] D. OBrien, S. Biswas, S. M. Imtiaz, R. Abdalkareem, E. Shihab, and H. Rajan, "Are prompt engineering and todo comments friends or foes? an evaluation on github copilot," in *Proceedings of the IEEE/ACM 46th International Conference on Software Engineering (ICSE)*, 2024, pp. 1–13.

[9] A. Mastropaolo, L. Pascarella, E. Guglielmi, M. Ciniselli, S. Scalabrino, R. Oliveto, and G. Bavota, "On the robustness of code generation techniques: An empirical study on github copilot," in *IEEE/ACM 45th International Conference on Software Engineering (ICSE)*, 2023, pp. 2149–2160.

[10] H. Pearce, B. Ahmad, B. Tan, B. Dolan-Gavitt, and R. Karri, "Asleep at the keyboard? assessing the security of github copilot's code contributions," in *IEEE Symposium on Security and Privacy (S&P)*, 2022, pp. 754–768.

[11] J. Y. Khan and G. Uddin, "Automatic code documentation generation using gpt-3," in *Proceedings of the 37th IEEE/ACM International Conference on Automated Software Engineering (ASE)*, 2022, pp. 1–6.

[12] K. Li, Q. Hu, J. Zhao, H. Chen, Y. Xie, T. Liu, M. Shieh, and J. He, "Instructcoder: Instruction tuning large language models for code editing," in *Proceedings of the 62nd Annual Meeting of the Association for Computational Linguistics (Volume 4: Student Research Workshop)*, 2024, pp. 50–70.

[13] R. Tufano, S. Masiero, A. Mastropaolo, L. Pascarella, D. Poshyvanyk, and G. Bavota, "Using pre-trained models to boost code review automation," in *Proceedings of the 44th International Conference on Software Engineering (ICSE)*, 2022, pp. 2291–2302.

[14] A. M. Dakhel, V. Majdinasab, A. Nikanjam, F. Khomh, M. C. Desmarais, and Z. M. J. Jiang, "Github copilot ai pair programmer: Asset or liability?" *Journal of Systems and Software (JSS)*, vol. 203, p. 111734, 2023.

[15] C. Dantas, A. Rocha, and M. Maia, "Assessing the readability of chatgpt code snippet recommendations: A comparative study," in *Proceedings of the XXXVII Brazilian Symposium on Software Engineering(SBES)*, 2023, p. 283–292.

[16] Z. Liu, Y. Tang, X. Luo, Y. Zhou, and L. F. Zhang, "No need to lift a finger anymore? assessing the quality of code generation by chatgpt," *IEEE Transactions on Software Engineering (TSE)*, 2024.

[17] D. Nam, A. Macvean, V. Hellendoorn, B. Vasilescu, and B. Myers, "Using an llm to help with code understanding," in *Proceedings of the IEEE/ACM 46th International Conference on Software Engineering (ICSE)*, 2024, pp. 1–13.

[18] J. Choi, G. An, and S. Yoo, "Iterative refactoring of real-world open-source programs with large language models," in *International Symposium on Search Based Software Engineering*. Springer, 2024, pp. 49–55.

[19] Z. Ji, N. Lee, R. Frieske, T. Yu, D. Su, Y. Xu, E. Ishii, Y. J. Bang, A. Madotto, and P. Fung, "Survey of hallucination in natural language generation," *ACM Computing Surveys*, vol. 55, no. 12, pp. 1–38, 2023.

[20] O. Asare, M. Nagappan, and N. Asokan, "Is github's copilot as bad as humans at introducing vulnerabilities in code?" *Empirical Software Engineering (EMSE)*, vol. 28, no. 6, p. 129, 2023.

[21] N. Nguyen and S. Nadi, "An empirical evaluation of github copilot's code suggestions," in *Proceedings of the 19th International Conference on Mining Software Repositories (MSR)*, 2022, p. 1–5.

[22] R. I. T. Jensen, V. Tawosi, and S. Alamir, "Software vulnerability and functionality assessment using llms," in *2024 IEEE/ACM International Workshop on Natural Language-Based Software Engineering (NLBSE)*. IEEE, 2024, pp. 25–28.

[23] J. Zhu, Y. Miao, T. Xu, J. Zhu, and X. Sun, "On the effectiveness of large language models in statement-level code summarization," in *2024 IEEE 24th International Conference on Software Quality, Reliability and Security (QRS)*. IEEE, 2024, pp. 216–227.

[24] D. Pomian, A. Bellur, M. Dilhara, Z. Kurbatova, E. Bogomolov, A. Sokolov, T. Bryksin, and D. Dig, "Em-assist: Safe automated extractmethod refactoring with llms," in *Companion Proceedings of the 32nd ACM International Conference on the Foundations of Software Engineering*,

[25] A. Shirafuji, Y. Oda, J. Suzuki, M. Morishita, and Y. Watanobe, "Refactoring programs using large language models with few-shot examples," in *2023 30th Asia-Pacific Software Engineering Conference (APSEC)*. IEEE, 2023, pp. 151–160.

[26] "What is a code smell?" https://www.sonarsource.com/learn/code-smells/, 2024, access: Oct 2nd, 2024.

[27] Z. Zeng, Y. Wang, R. Xie, W. Ye, and S. Zhang, "Coderujb: An executable and unified java benchmark for practical programming scenarios," in *Proceedings of the 33rd ACM SIGSOFT International Symposium on Software Testing and Analysis*, 2024, pp. 124–136.

[28] C.-Y. Su, A. Bansal, V. Jain, S. Ghanavati, and C. McMillan, "A language model of java methods with train/test deduplication," in *Proceedings of the 31st ACM Joint European Software Engineering Conference and Symposium on the Foundations of Software Engineering*, 2023, pp. 2152–2156.

[29] "The top programming languages," https://octoverse.github.com/2022/top-programming-languages, 2024, access: Oct 2nd, 2024.

[30] F. Hassan, S. Mostafa, E. S. Lam, and X. Wang, "Automatic building of java projects in software repositories: A study on feasibility and challenges," in *ACM/IEEE International Symposium on Empirical Software Engineering and Measurement (ESEM)*, 2017, pp. 38–47.

[31] D. Marcilio, R. Bonifácio, E. Monteiro, E. Canedo, W. Luz, and G. Pinto, "Are static analysis violations really fixed? a closer look at realistic usage of sonarqube," in *IEEE/ACM 27th International Conference on Program Comprehension (ICPC)*, 2019, pp. 209–219.

[32] C. Vassallo, S. Panichella, F. Palomba, S. Proksch, A. Zaidman, and H. C. Gall, "Context is king: The developer perspective on the usage of static analysis tools," in *IEEE 25th International Conference on Software Analysis, Evolution and Reengineering (SANER)*, 2018, pp. 38–49.

[33] C. Vassallo, F. Palomba, A. Bacchelli, and H. C. Gall, "Continuous code quality: Are we (really) doing that?" in *Proceedings of the 33rd ACM/IEEE International Conference on Automated Software Engineering (ASE)*, 2018, pp. 790–795.

[34] Y. Wang, Q. Yao, J. T. Kwok, and L. M. Ni, "Generalizing from a few examples: A survey on few-shot learning," *ACM computing surveys (csur)*, vol. 53, no. 3, pp. 1–34, 2020.

[35] S. Gao, X.-C. Wen, C. Gao, W. Wang, H. Zhang, and M. R. Lyu, "What makes good in-context demonstrations for code intelligence tasks with llms?" in *2023 38th IEEE/ACM International Conference on Automated Software Engineering (ASE)*. IEEE, 2023, pp. 761–773.

[36] K. Constantino, F. Belem, and E. Figueiredo, "Dual analysis for helping developers to find collaborators based on co-changed files: An empirical study," *Software: Practice and Experience (SPE)*, 2023.

[37] J.-Y. Yao, K.-P. Ning, Z.-H. Liu, M.-N. Ning, and L. Yuan, "Llm lies: Hallucinations are not bugs, but features as adversarial examples," *arXiv preprint arXiv:2310.01469*, 2023.

[38] B. A. Muse, F. Khomh, and G. Antoniol, "Refactoring practices in the context of data-intensive systems," *Empirical Software Engineering*, vol. 28, no. 2, p. 46, 2023.

[39] S. Shafiq, W. K. Assunção, A. Mashkoor, C. Mayr-Dorn, and A. Egyed, "Towards recommending refactoring operations based on bugs," *Available at SSRN 4397230*, 2024.

[40] G. Vale, C. Hunsen, E. Figueiredo, and S. Apel, "Challenges of resolving merge conflicts: A mining and survey study," *IEEE Transactions on Software Engineering (TSE)*, 2021.

[41] Jetbrains, "Jetbrains - code refactoring," https://www.jetbrains.com/help/phpstorm/extract-constant.html, 2024, access: Oct 2nd, 2024.

[42] Z. Xian, C. Cui, R. Huang, C. Fang, and Z. Chen, "zs-llmcode: An effective approach for functional code embedding via llm with zero-shot learning," *arXiv preprint arXiv:2409.14644*, 2024.

[43] R. Inoue and Y. Higo, "Improving accuracy of llm-based code clone detection u sing functionally equivalent methods," in *2024 IEEE/ACIS 22nd International Conference on Software Engineering Research, Management and Applications (SERA)*. IEEE, 2024, pp. 24–27.

[44] G. Santos, A. Santana, G. Vale, and E. Figueiredo, "Yet another model! a study on model's similarities for defect and code smells," in *Proceedings of the 26th International Conference on Fundamental Approaches to Software Engineering (FASE)*, 2023.

[45] Y. Qin, S. Wang, Y. Lou, J. Dong, K. Wang, X. Li, and X. Mao, "Agentfl: Scaling llm-based fault localization to project-level context," *arXiv preprint arXiv:2403.16362*, 2024.

[46] J. Li, A. Sangalay, C. Cheng, Y. Tian, and J. Yang, "Fine tuning large language model for secure code generation," in *Proceedings of the 2024 IEEE/ACM First International Conference on AI Foundation Models and Software Engineering*, 2024, pp. 86–90.

[47] A. Z. Yang, C. Le Goues, R. Martins, and V. Hellendoorn, "Large language models for test-free fault localization," in *Proceedings of the 46th IEEE/ACM International Conference on Software Engineering (ICSE)*, 2024, pp. 1–12.

[48] Z. Ma, A. R. Chen, D. J. Kim, T.-H. P. Chen, and S. Wang, "Llmparser: An exploratory study on using large language models for log parsing," in *Proceedings of the 46th IEEE/ACM International Conference on Software Engineering (ICSE)*, 2024.

[49] J. Hoffmann and D. Frister, "Generating software tests for mobile applications using fine-tuned large language models," in *Proceedings of the 5th ACM/IEEE International Conference on Automation of Software Test (AST)*, 2024, pp. 76–77.